\documentclass[12pt]{article}

\usepackage[english]{babel}
\usepackage{pazh}
\usepackage{graphicx}
\tightenlines

\sloppypar

\begin{document}

\title{\bf Supernovae -- Optical Precursors of Short Gamma-Ray Bursts}

\author{\bf \hspace{-1.3cm} \ \
V.I. Dokuchaev, Yu.N. Eroshenko
}

\affil{\it Institute for Nuclear Research, Russian Academy of Sciences, 
pr. 60-letiya Oktyabrya 7a, Moscow, 117319 Russia}

\sloppypar \vspace{2mm} \noindent 
The probability of observing ``supernova -- gamma-ray burst'' (GRB) pair events and recurrent
GRBs from one galaxy in a time interval of several years has been estimated. Supernova explosions in
binary systems accompanied by the formation of a short-lived pair of compact objects can be the sources
of such events. If a short GRB is generated during the collision of a pair, then approximately each of $\sim300$ short GRBs with redshift $z$ must have an optical precursor -- a supernova in the observer's time interval $<2\cdot(1+z)$~yr. If the supernova explosion has the pattern of a hypernova, then a successive observation of long and short GRBs is possible. The scenario for the generation of multiple GRBs in collapsing galactic nuclei is also discussed.

\section*{INTRODUCTION}

Progress in investigating gamma-ray bursts
(GRBs) in the last decade is associated primarily
with the observation of their radio, optical, and X-ray
afterglows. Concurrently, the theoretical models
of stellar evolution and supernova explosions, in
particular, the models for the collimation of high-energy
jets probably responsible for the generation
of GRBs, are improved. According to the most
popular scenarios, long GRBs (with a duration of
more than 2~s) emerge during supernova explosions
called hypernovae (Paczynski 1998; MacFadyen and
Woosley 1999), while short bursts are produced
during the coalescence of two neutron stars (NSs)
or a NS and a black hole (BH) in binary systems
(Blinnikov et al. 1984). However, the models in
which long GRBs can be generated at the final stage
of the collision between compact and ordinary stars in
a binary system have also been proposed (Barkov and
Komissarov 2009).

GRBs that have a complex structure with a long
time interval between gamma-ray peaks and a delay
between gamma-ray, optical, and X-ray maxima
deserve special attention. A significant duration of
the light curves for some GRBs (up to $2\cdot10^4$~s)
and the presence of precursor peaks suggest that
the central engine can operate much longer than
the gamma-ray generation stage (Lipunov and Gorbovskoy
2007). Apart from bursts with a complex
structure, several GRBs that can be considered as recurrent ones or as clusters of bursts may have been
observed. One of the most unusual events of this kind
was recorded in October 1996, when four separate
GRBs came from a small region of the sky during
two days (Meegan et al. 1996; Graziani et al. 1998).
The probability of a chance projection of the sources
of independent GRBs onto the same line of sight is
only $3.1\cdot10^{-5}\div3.3\cdot10^{-4}$. Since in an individual
galaxy the NS-NS or NS-BH coalescences in binary
systems occur, on average, once in $10^5-10^6$~yr,
the generation of four GRBs in one galaxy during
two days seems highly unlikely in this model. At
the same time, the spectra of these GRBs rule out
their interpretation as the emission of a soft gamma
repeater. However, this event can be attributed to the
class of superlong (with a duration $>500$~s) GRBs;
several such bursts were detected in the BATSE catalog
(Tikhomirova and Stern 2005). If the events
recorded in October 1996 belong to different episodes
of superlong bursts, then the probability that these
events are recurrent GRBs is very low (Graziani et al.
1998; Tikhomirova and Stern 2005). GRB clusters
were also searched for previously by the methods of
statistical analysis (Kuznetsov 2001).

Several multistage supernova core collapse models
and explanations of GRBs with a complex structure
have been proposed. The model for the fragmentation
of a rotating supernova core with the formation
of several low-mass NSs or BHs that subsequently
coalesce under the influence of losses through
gravitational radiation was discussed by Imshennik
and Nadyozhin (1992) and Imshennik (1995). In particular, the possibly observed double neutrino burst from
SN~1987A (Stella and Treves 1987; Berezinsky et al.
1988; Imshennik and Nadyozhin 1988; Imshennik
and Ryazhskaya 2004) can be explained in such models.
Davies et al. (2002) investigated the scenario in
which the fragments coalesce several hours or days
after the supernova explosion and a GRB is generated
during their collision. The intense peaks in
the GRB profile (e.g., in GRB~050502b) long after
its onset can be explained in a similar model (King
et al. 2005). Variants with two collapse stages with
the formation of initially a NS and subsequently a
BH have been proposed (Rujula 1987; Hillebrandt
et al. 1987). Successive high-energy events in a
short time interval are also possible in this model.
GRBs with a precursor in a time interval of $\sim100$~s
can be explained in the spinar model (Lipunov and
Gorbovskoy 2007; Lipunova et al. 2009). Superlong
GRBs (with a duration up to several thousand
seconds) and precursor peaks in the GRB time profile
can emerge at the long collapse stage of the coalesced
stars in a binary system (Barkov and Komissarov
2009). The observation of multiple GRBs can
also be the result of gravitational lensing (Babichev
and Dokuchaev 2000).

In this paper, we investigate the limiting case of
the standard picture (Hills 1983; Grishchuk et al.
2001) of supernova explosions in binary systems
where a fully-fledged (unfragmented) NS or BH
is produced by the explosion that acquires a kick
velocity, passes to a new orbit, and then, under
gravitational radiation, comes closer to and coalesces
with the second compact object of the pair. A change
in parameters is possible during the explosion when
the coalescence occurs only a few years after the
supernova explosion. Thus, the generation of a
recurrent high-energy event from one source in a
short time interval is also possible within the standard
picture of a supernova explosion. The main goal of
this paper is to estimate the probability of such events
and to discuss the prospects for their recording.

As an alternative scenario for the generation of
multiple GRBs, we briefly discuss the scenario of collisions
between NSs and BHs in collapsing galactic
nuclei. These collisions can serve as GRB sources.
The dynamical evolution of a NS cluster can lead to
its gravitational collapse in the process of avalanche-like
contraction (Zel'dovich and Podurets 1965;
Shapiro and Teukolsky 1986). Multiple GRBs can
be generated during this stage in a time interval of
several days. The GRB detected in October 1996
(Meegan et al. 1996; Graziani et al. 1998) can be
naturally explained in this model.

\section*{COLLISIONS IN BINARY SYSTEMS}

In this section, we will follow the notation from
Grishchuk et al. (2001), which presents the main
results of the kinematics of binary star systems and
their evolution as a result of supernova explosions
and gravitational wave emission. Additionally, we will
take into account the gravitational capture effect.

Consider a binary system in a circular orbit that
consists of a star with mass $M_1$ and a compact object
(NS or BH) with mass $M_2$. In reality, the orbits
of binary systems with compact objects are highly
eccentric and the supernova explosion can occur at
any point of their elliptical orbits. It can be assumed
that the explosion is most likely at apastron and periastron,
where the star undergoes gravitational kicks
from tidal forces. Gravitational kicks can serve as
triggers of the explosion of an unstable star on the
verge of collapse, but this effect has not yet been
investigated. Here, we consider a circular orbit to
obtain an analytical estimate. Although this approach
is not quite accurate, it simplifies considerably the
calculations and, in a sense, serves as an averaging of
the actual elliptical orbit. A more accurate calculation
could be performed by population synthesis methods.

The star $M_1$ explodes as a supernova, leaving a
compact remnant with mass $M_c$ behind. We will
mark the quantities immediately before and after the
explosion by the subscripts ``i'' and ``f'', respectively.
Denote $M_i=M_1+M_2$, $M_f=M_c+M_2$, and $\mu_f=M_cM_2/(M_c+M_2)$. The initial separation between
the pair components is $a_i$ and the initial relative velocity
is $v_i=(GM_i/a_i)^{1/2}$. We will direct the x axis
from $M_2$ to $M_1$ and the y axis along $\vec v_i$. After the
explosion, $M_c$ acquires a kick velocity $\vec w$ in the rest
frame of $M_1$ in such a way that $\vec v_f=(w_x,v_i+w_y,w_z)$.
The orbital parameters after the explosion can be
found from the energy and angular momentum conservation
laws. We will write the energy conservation
law as
\begin{equation}
\frac{\mu_fv_f^2}{2}-\frac{GM_cM_2}{a_i}-\Delta E_{\rm
gw}=-\frac{GM_cM_2}{2a_f}, \label{econs}
\end{equation}
where, in contrast to Grishchuk et al. (2001), we
added the energy losses through gravitational radiation
during the first approach $\Delta E_{\rm
gw}$.

\begin{figure}[t]
\includegraphics[width=0.95\textwidth]{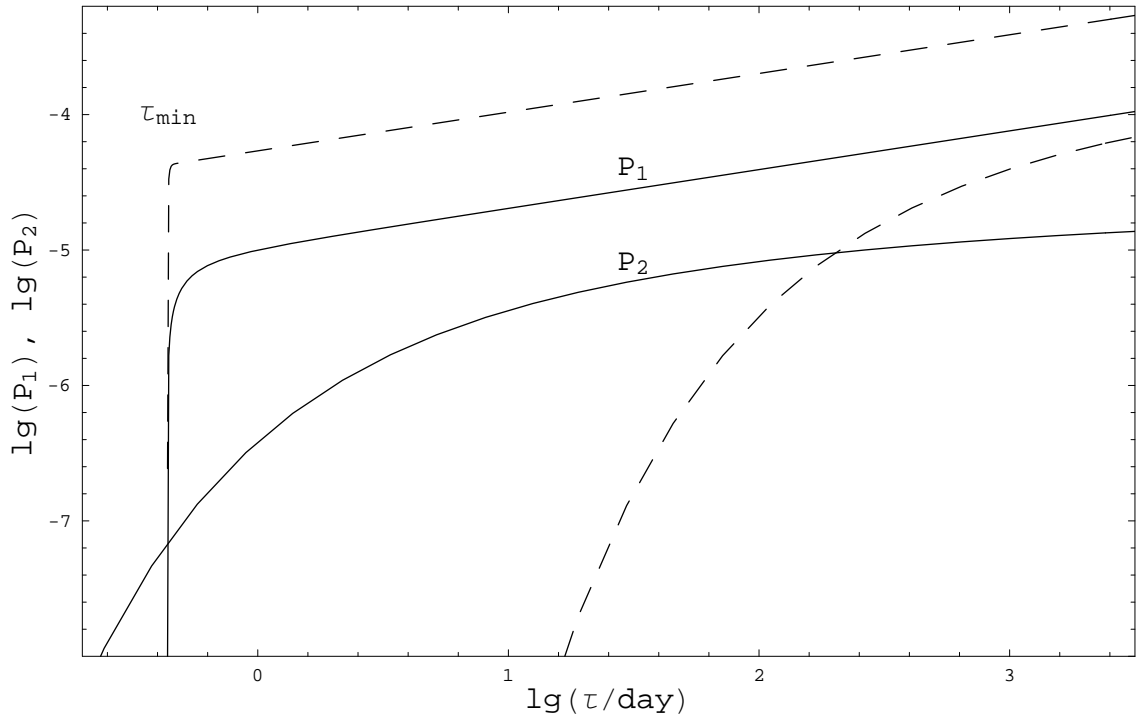}
\caption{The two upper curves indicate the formation probability $P_1(<\tau)$ of a pair of compact objects with a lifetime $<\tau$ (3). The
lower curves indicate the probability $P_2(<\tau)$ in the case of a gravitational capture (9). The solid curves correspond to an
isotropic kick velocity distribution $\sigma_x=\sigma_y=\sigma_z=250$~km~s$^{-1}$ and the dashed curves correspond to $\sigma_x=\sigma_z=50$~km~s$^{-1}$
and $\sigma_y=250$~km~s$^{-1}$.} \label{fig1}
\end{figure}

In most situations, the gravitational radiation
is very weak and the term $\Delta E_{\rm gw}$ in (1) may be
neglected. Let us first consider this case. Given
the angular momentum conservation law, well known
results for the new orbital parameters are
then obtained (Grishchuk et al. 2001). The binary
system remains bound under the condition
$v_f<v_i(2/\chi)^{1/2}$, where $\chi=M_i/M_f$. Subsequently,
slow secular contraction of the binary through the
emission of gravitational waves takes place. We
are interested in the case where a short-lived binary
whose orbit has a high ellipticity $1-e_f\ll 1$ is formed
after the explosion. From Eqs. (A.22)-(A.24) in
Grishchuk et al. (2001), we derive an asymptotic
(when $1-e_f\ll 1$) expression for the lifetime of the pair:
\begin{equation}
t_f=\frac{3c^5a_f^4(1-e_f^2)^{7/2}}{85G^3M_f^2\mu_f}. \label{tf1}
\end{equation}
Let us require that the lifetime be less than some
fixed value: $t_f<\tau$. Below, we will assume that the
distribution in velocity $\vec w$ is Gaussian (Grishchuk
et al. 2001) in each component with the root-means-square values of $\sigma_x$, $\sigma_y$, and $\sigma_z$. Note that
other (non-Gaussian) distributions fitting the pulsar
velocity data were also proposed (for a discussion,
see, e.g., Grishchuk et al. 2001). The condition
$t_f<\tau$ separates out the region corresponding to
the event we consider in velocity space. Integrating
the Gaussian distribution over the velocity
components $w_x$, $w_y$, and $w_z$ under the condition
$t_f<\tau$, we obtain the probability of the necessary
change in orbit:
\begin{equation}
 P_1(t_f<\tau)\simeq\frac{v_i^3A^{2/7}}{2\sqrt{2\pi}\sigma_x
 \sigma_y\sigma_z}\,e^{-v_i^2/2\sigma_y^2}\int\limits_0^{p_{\rm max}}\!dp
 \,(2-\chi p^2)^{1/7}e^{-p^2v_i^2/2\sigma_x^2},
 \label{p1}
\end{equation}
where
\begin{equation}
A=\frac{85M_f^2\mu_fv_i^8}{3c^5GM_i^4\chi^{7/2}}\tau,
\end{equation}
and $p_{\rm max}$ corresponds to the condition that the period
of the new orbit does not exceed $\tau$:
\begin{equation}
p_{\rm max}^2=\frac{2}{\chi}-\left(\frac{2\pi
GM_f}{v_i^3\tau}\right)^{2/3}.
\end{equation}
At low $\tau$, the probability $P_1(t_f<\tau)$ may turn out
to be zero, because not only the time of contraction
through gravitational radiation but also even the orbital
period $T_f$ of the new orbit will be greater than $\tau$.
From the condition $p_{\rm max}=0$, we obtain the minimum
possible time $\tau$:
\begin{equation}
\tau_{\rm min}=\frac{2\pi
GM_f}{v_i^3}\left(\frac{\chi}{2}\right)^{3/2}.
\end{equation}
In (3), we disregarded the low probability of a direct
hit at exact compensation $w_y=-v_i$, $w_z=0$. However,
this case was included in the probability of a
direct capture calculated in the next section.

As an example, let us perform calculations for
the following typical set of parameters: 
$M_1=15M_{\odot}$, $M_2=10M_{\odot}$,
$M_c=1.4M_{\odot}$, $T_i=1$~day,
$\sigma_x=\sigma_y=\sigma_z=250$~km~s$^{-1}$; then,$\tau_{\rm
min}=0.44$~day. The derived
probability (3) is very low (see the figure). We are
interested in the ratio of quantity (3) to another small
quantity $P_{\rm tot}$ -- the probability that a binary with a
lifetime less than the age of galaxies will be produced
by the explosion. To accurately calculate $P_{\rm tot}$, we need
detailed population synthesis calculations; it is also
necessary to take into account the distribution of binary
systems in their component masses and in $a_i$ as
well as the gas-dynamic interaction between the stars.
However, no such detailed calculation is required for
the purposes of this paper. The result (3) becomes
inaccurate at large $\tau$, because it was obtained in the
limit $e_f\to 1$, but as an acceptable estimate we can
assume that $P_1(t_f<\tau)\propto \tau^{2/7}$. The fraction of the
cases where the collision occurs a time $<\tau$ after the
supernova explosion can then be found as follows:
\begin{equation}
P_{\rm rel}=P_1/P_{\rm tot}\sim(\tau/t_G)^{2/7}\simeq3\cdot10^{-3}
\left(\frac{\tau}{2\mbox{~yr}}\right)^{2/7}, \label{prelmain}
\end{equation}
where $t_G\simeq12\cdot10^9$~yr is the characteristic age of the
galaxies formed at $z_G\simeq4$. The relative probability (7)
is the main result of this paper. A slightly lower value,
$P_1(t_f<2\mbox{~yr})/P_1(t_f<t_G)\sim1.6\cdot10^{-3}$, is obtained
in a direct calculation based on Eq. (3) for the above
set of parameters.

Thus, if the collisions of compact objects in pairs
produce short GRBs, then approximately one GRB
preceded by a supernova explosion in the observer's
time interval $<2\cdot(1+z)$~yr, where $z$ is the redshift of
the GRB source, will be found among every $\simeq300$
such GRBs. In addition to the optical afterglow, such
GRBs have an optical precursor. Since several thousand
GRBs have been detected to date, the events
considered here can be among them.

Long GRBs can also be generated during supernova
explosions if the explosion has the pattern
of a hypernova (Paczynski 1998; MacFadyen and
Woosley 1999). If a hypernova exploded in a binary
system and if a collision involving a NS subsequently
occurred after time $\tau$, then we have two successive
GRBs, long and short ones. However, since only
a small fraction of collapsing massive stars produce
an observable long GRB (the GRB is seen only if
the collimated jet is directed toward the observer), the
probability of observing recurrent GRBs is very low
and, to all appearances, is outside the observational
capabilities.

From the viewpoint of observations, it is also necessary
to estimate the probability of a chance projection
of an independent supernova into the error box of
the observed GRB. The question about the properties
of such GRBs is also of considerable interest, because
the collision of compact objects occurs in the medium
that left after a recent supernova explosion and that
affects the expanding fireball.

To ascertain how reasonable the values given by
Eq.~(3) are at large $\tau$, let us make the following simple
estimate. The rate of supernova explosions in one
galaxy is $\sim2\cdot10^{-2}$~yr$^{-1}$. About 50\% of the stars are
members of binary systems. Let us assume that the
pair survival probability during the first explosion is
$\sim0.5$. It follows from Eq.~(3) for our set of typical parameters
that
$P_1(t_f<t_G)\sim4\cdot10^{-2}$. Multiplying these numbers
yields an estimate of the collision rate per galaxy
$\sim2\cdot10^{-4}$~yr$^{-1}$, which is consistent, in order of magnitude,
with the population synthesis calculations. A
modification of the scenario considered in this section
could be a consideration of the change in the orbits
of compact objects during supernova explosions in
triple hierarchical systems, but the probability of the
corresponding events is much lower.

The kick velocity acquired by a compact supernova
remnant has a significant effect on the collision rate
of binary NSs and BHs (see Grishchuk et al. 2001;
Postnov and Kuranov 2008; Kuranov et al. 2009;
and references therein). Both the origin of the kick
velocity and its distribution in directions remain unclear
so far. The observed correlation between the velocity direction and the spin axis in young single
radio pulsars may be the key to understanding the
mechanism for the appearance of the kick velocity.
However, this correlation is not absolute and about
half of the pulsars have a large angle between these
directions. Variants with a non-central kick, when the
NS acquired an additional angular momentum during
the explosion, are also possible.

A rotating presupernova has the preferential directions
along the spin axis in which the kick velocity
may be directed, while the choice between these two
directions is determined by the inclination of the spin
axis to the direction of the orbital angular momentum
if the kick velocity is affected by the interaction
between the pair components (mass transfer, tidal
forces, or magnetic field structure). Another preferential
direction is the line connecting the two stars,
because it is along this line that the star is deformed
by the tidal gravitational forces and the explosion
nonsphericity may correspond to this direction. This
deformation can lead to an asymmetric, in the x direction,
explosion and, accordingly, to the appearance
of a kick velocity along the x axis. The combined
effect of various factors in such a way that the kick
velocity has different directions in binary systems with
different parameters is not ruled out either.

We performed calculations for several relations
between $\sigma_x$,
$\sigma_y$, and $\sigma_z$. In the case of $\sigma_x=\sigma_z=50$~km~s$^{-1}$ and $\sigma_y=250$~km~s$^{-1}$, the probability $P_1$
is higher than that for the isotropic case of $\sigma_x=\sigma_y=\sigma_z=250$~km~s$^{-1}$, because a lower kick velocity
along the x and z axes facilitates the compensation
of the orbital velocity along the y axis. However,
the case with the greatest $\sigma_y$ probably cannot be
realized, because the y direction is not preferential
in the above sense. On the contrary, in the cases of
$\sigma_x=250$~km~s$^{-1}$,
$\sigma_y=\sigma_z=50$~km~s$^{-1}$, and $\sigma_x=\sigma_y=50$~km~s$^{-1}$, $\sigma_z=250$~km~s$^{-1}$, the derived probability
is considerably lower (it is not shown in the
figure), because the compensation of the velocity $v_i$
is suppressed exponentially.

Thus, the kick velocity is fundamentally important
in determining the total probability. However, it
should be noted that our calculations of the relative
fraction of short-lived pairs are largely independent
of the pattern of the kick velocity, because it affects
both $P_1$ and $P_{\rm tot}$ in a similar way, while their ratio and
the universal dependence $P_{\rm
rel}\propto\tau^{2/7}$ remain almost
unchanged.

\section*{GRAVITATIONAL CAPTURE IN A BINARY
SYSTEM}

Let us now take into account the gravitational
capture effect, when the binary components pass
fairly close to each other after the supernova explosion
and an energy comparable to the total kinetic
energy of the binary components is emitted through
gravitational waves already during the first approach.
Using the results from Quinlan and Shapiro (1987),
we will obtain the energy emitted in the time $T_f/2$ (half of the orbital period),
\begin{equation}
\Delta E_{\rm gw}\simeq\frac{85\pi
G^{7/2}M_c^2M_2^2M_f^{1/2}}{3\cdot2^{7/2}c^5r_p^{7/2}},
\label{egv}
\end{equation}
where $r_p=a_f(1-e_f)$ is the minimum distance of the
closest approach during the first flyby. An exact
solution of the problem including strong gravitational
wave emission presents great difficulties. However,
the following simple estimate can be made. During
the gravitational capture, an energy comparable to
the kinetic energy is emitted through gravitational
waves. In this case, the first and third terms on the
left-hand side of Eq.~(1) and the term on the right-hand
side are all of the same order of magnitude.
Hence, integrating, as in Eq.~(3), over the distribution
of velocities $\vec w$, we obtain
\begin{equation}
 P_2(t_f<\tau)\simeq\frac{v_i^3B^{2/7}}
 {2\sqrt{2\pi}\sigma_x\sigma_y\sigma_z}\,e^{-v_i^2/2
 \sigma_y^2}\int\limits^{\infty}_{p_{\rm min}
 (\tau)}\!\!dp\,p^{-4/7}e^{-p^2v_i^2/2\sigma_x^2},
 \label{p2}
\end{equation}
where
\begin{equation}
B=\frac{85\pi M_cM_2v_i^5}{3c^5M_i^2\chi^5},
\end{equation}
\begin{equation}
 p_{\rm min}(\tau)={\rm max}\left\{\frac{GM_i}{v_i^3\tau};
 ~~\left(\frac{\sqrt{2}\pi GM_f}{v_i^3\tau}\right)^{1/3}\right\}.
 \label{pmin}
\end{equation}
Here, the first term in parentheses corresponds to
the condition that $M_c$ will reach $M_2$ in time $\tau$, while
the second term was obtained from the relation $t_f=\tau$, where $t_f$ is expressed by (2). If all masses are
of the same order of magnitude, then $B/A\sim T_i/\tau$,
i.e., the gravitational capture effect dominates only for
very low $\tau$, less than the initial orbital period. The
results of our calculations in the case of gravitational
captures are shown in the figure. At $\tau\ge0.5$~day, the
gravitational capture effect may be neglected.

Note that the binary system can remain gravitationally
bound even if $v_f>v_i(2/\chi)^{1/2}$ due to the gravitational
capture. Thus, including the gravitational
capture removes the constraint on $v_f$, although the
capture probability is very low.

\section*{RECURRENT GAMMA-RAY BURSTS
FROM GALACTIC NUCLEI}

In this section, we will discuss an alternative GRB
recurrence scenario that associates recurrent and
multiple ($>2$ events) bursts with the final evolutionary
stage of the central star clusters in galactic nuclei
and the formation of supermassive BHs. Bursts are
generated during random collisions between NSs
and BHs in clusters. Some elements of this
model were discussed previously (Dokuchaev et al.
1997, 1998).

Dense star clusters are present in the nuclei of
most large structured galaxies. For spiral galaxies,
the masses of these clusters correlate with
the total stellar masses of the galaxies (Erwin and
Gadotti 2010). A compact cluster of NSs and BHs
must be formed at the center of an initial cluster of
ordinary stars through mass segregation (the settling
of more massive stars to the cluster center) and
supernova explosions (Colgate 1967; Sanders 1970).
Below, it is this inner subsystem of compact objects
that we call a cluster. Stellar-mass BHs are formed
during pair collisions of NSs that subsequently continue
to build up their mass through recurrent mutual
coalescences. Concurrently, the central region of
the cluster contracts through the processes of pair
relaxation, which can result in the loss of stability by
it and gravitational collapse into a supermassive BH.
Although most supermassive BHs in galactic nuclei
have probably been formed according to a different
scenario, through the collapses of gaseous clouds
in protogalaxies (for a review, see Dokuchaev et al.
2007), the model of a collapsing cluster is, in a sense,
an inevitable evolutionary stage of galactic nuclei.
The possibility of a gravitational collapse is limited
only by the time needed for pair relaxation of stars
in the cluster (Dokuchaev 1991). In this section,
we will assume that the gravitational collapses of
central clusters consisting of stellar-mass BHs with
an admixture of non-coalesced Ns actually took place
in a significant fraction of galaxies, $f_c\sim
1$. We
consider the fraction of NSs in the cluster fNS at
the stage immediately before its collapse to be a free
parameter that depends on the cluster's evolutionary
track.

Single GRBs can be generated in clusters that
are still far from the collapse stage (Dokuchaev et al.
1998). However, here we are interested in the cases
of fast GRB recurrence. Such GRBs can be generated
directly in the process of cluster core collapse
accompanied by rapid avalanche-like contraction at
the cluster center (Zel'dovich and Podurets 1965;
Shapiro and Teukolsky 1986). We will show that
the avalanche can last $\sim 2$~days when some plausible
cluster parameters are chosen and about four NS coalescences
will occur in this time, which corresponds
to the multiple burst detected on October 27, 1996
(Meegan et al. 1996; Graziani et al. 1998).

The BH masses in the central part of a collapsing
cluster grow significantly by the time of its collapse
through coalescences (Quinlan and Shapiro 1987).
Suppose that their masses immediately before the
collapse are $m_{\rm BH}\sim10m$, where $m\simeq1.4M_{\odot}$ is the
NS mass. Suppose also that the cluster has mass $M$
(predominantly in the form of BHs) and that NSs
account for a small fraction $f_{\rm
NS}\ll1$ of this mass.
The effective radius of the cluster is $R=MG/2v^2$,
where $v$ is the velocity dispersion. When the cluster
reaches the collapse stage, a dense core with
mass $M_c$, radius $R\simeq3R_{g,c}$, where $R_{g,c}=2GM/c^2$,
and velocity dispersion $u\simeq0.3c$ (at the marginally
stable orbit) emerges at its center. All stars with
an angular momentum $J<2mcR_{g,c}$ will fall along
a spiral to the core without returning to the cluster.
The presence of quasi-elliptical orbits that connect
different layers of stars between themselves are of
crucial importance for the growth of the avalanche
of falling stars (Zel'dovich and Podurets 1965). According
to the numerical calculations by Shapiro and
Teukolsky (1986), several percent of the total cluster
mass will fall onto the forming BHin a dynamical time
$t_{\rm dyn}=R/v$.

At $f_{\rm NS}\ll1$, the NS-BH coalescences are more
efficient than the NS-NS ones. At the above
mass ratio $m_{\rm BH}\sim10m$, the NS does not vanish
under the BH horizon as a whole but is disrupted
by tidal forces to produce a relativistic fireball and
to generate a GRB. The rate of NS-BH coalescences
in the core can be calculated from the formula
$\dot N_c=\sigma_{\rm cap}uN_{\rm
NS}n_{\rm BH}$, where the gravitational
capture cross section $\sigma_{\rm
cap}$ is given in Mouri and
Taniguchi (2002), $N_{\rm NS}=f_{\rm
NS}M_c/m$ is the number
of NSs in the core, and $n_{\rm BH}$ is the BH number
density in the core. If we choose $M=10^7M_{\odot}$ and
$v=0.066c$, then the cluster's dynamical time (core
collapse duration) will be $t_{\rm
dyn}=GM/v^3=2$~days.
The number of NS-BH collisions in the core in this
time is
\begin{equation}
\dot N_c\, t_{\rm dyn}\simeq4 \left(\frac{f_{\rm
NS}}{10^{-4}}\right)
\left(\frac{M_c}{5\cdot10^5M_{\odot}}\right)^{-1}.
\end{equation}
The rate of NS-BH collisions in the entire cluster
before its collapse can be estimated similarly:
\begin{equation} \dot N\simeq7\cdot10^{-3}
\left(\frac{f_{\rm NS}}{10^{-4}}\right)
\left(\frac{M}{10^7M_{\odot}}\right)^{-1}
\left(\frac{v}{0.066c}\right)^{31/7} \mbox{yr}^{-1}, \label{e3}
\end{equation}
i.e., the cluster before its collapse was not the source
of bursts with fast recurrence. As a result, we have
four NS-BH collisions and, accordingly, four GRBs
from the collapsing core in a time of two days, which
is required to explain the temporal characteristics of
the multiple burst (Meegan et al. 1996; Graziani et al.
1998). Some influence on the properties of the fireball
in this model can be exerted by its scattering by the
forming central supermassive BH.

If the BH formation epoch lasts $t_0\simeq10$~Gyr, then
the total rate of such collapses in the observable Universe
is estimated as
\begin{equation}
\dot N_h\sim\frac{4\pi}{3}(ct_0)^3f_cn_gt_0^{-1}\approx0.1
\left(\frac{f_cn_g}{10^{-2}\mbox{Mpc}^{-3}}\right)
\mbox{yr}^{-1},
\end{equation}
where $n_g$ is the observed number density of structured
galaxies. Since the observations have been carried
out for almost 40 years, it seems natural that there
can be multiple GRBs predicted by this model among
the observed GRBs.

\section*{CONCLUSIONS}

We estimated the probability of supernova-GRB
pair events. One of every $\sim 300$ short GRBs in a
time interval of $\sim2$~yr was shown to have an optical
precursor -- the supernova during the explosion
of which the second compact object was born in the
pair. The detection of such supernova-GRB pairs
will serve as a weighty argument for the model of
the coalescence of binary compact objects as the
sources of short GRBs first proposed by Blinnikov
et al. (1984). We also considered two mechanisms
for the generation of recurrent GRBs from one
point of the sky in time intervals of several years
and several days. The first mechanism is related to
the evolution of stars in close binary systems. In
this model, the first GRBi s generated during a supernova
explosion, while the second GRB emerges
during the collision of compact stars -- the supernova
remnants. However, the probability of observing
such recurrent GRBs is very low, because only
a small fraction of exploding massive stars produce
a detectable GRB. In the second scenario, recurrent
GRBs are generated in evolved star clusters in
galactic nuclei immediately before the gravitational
collapse of their dense central regions. Since recurrent
GRBs are generated in the medium ``prepared''
by the preceding GRBs, the succeeding GRBs in
both scenarios can carry information about their generation
conditions in their spectral-temporal characteristics.
The evolution of multiple fireballs was
considered by Berezinsky and Dokuchaev (2006). Although
the events recorded on October 27, 1996,
cannot be attributed to short bursts, the evolution
of the fireball in the dense gaseous medium of the
galactic nucleus perturbed by the preceding fireballs
can be different from that in an isolated binary system.
Therefore, the question about the properties and
duration of the GRBs produced by the collisions of
compact objects in central clusters remains an open
one. The preceding (on long time scales) supernova
explosions or GRBs can also be revealed by the
observations of giant gaseous arcs and induced star
formation at the boundary of an expanding gaseous
bubble (Efremov et al. 1998). Supernova explosions
and the collisions of compact objects must be accompanied
by the emission of powerful signals in the
form of gravitational waves (Ruffert and Janka 1998;
Cheng and Wang 1999; Murphy et al. 2000; Grishchuk
et al. 2001). Although the searches for the
gravitational bursts accompanying GRBs have not
yielded any results (Abbott et al. 2009), gravitational
wave astronomy can provide decisive data on the
GRB mechanisms in the near future (Grishchuk et al.
2001).

\section*{ACKNOWLEDGMENTS}
We thank the referees of the paper for
many helpful remarks. This study was supported by
the Program of the Russian President for Support
of Scientific Schools (project no. NSh-3517.2010.2)
and the Federal Agency for Science and Innovations
(Rosnauka) (State contract no. 02.740.11.5092).

\section*{REFERENCES}

1. B. P. Abbott, R. Abbott, F. Acernese, et al.,
arXiv:0908.3824v1 [astro-ph.HE] (2009).

2. E. Babichev and V. Dokuchaev, Phys. Lett. A 265,
168 (2000).

3. M. V. Barkov and S. S. Komissarov,
arXiv:0908.0695v1 [astro-ph.HE] (2009).

4. V. S. Berezinsky and V. I. Dokuchaev, Astron. Astrophys.
454, 401 (2006).

5. V. S. Berezinsky, C. Castagnoli, V. I. Dokuchaev,
et al., Nuovo Cimento 11, 287 (1988).

6. S. I. Blinnikov, I. D. Novikov, T. V. Perevodchikova,
and A. G. Polnarev, Pis'ma Astron. Zh. 10, 422
(1984) [Sov. Astron. Lett. 10, 177 (1984)].

7. K. S. Cheng and J.-M.Wang, Astrophys. J. 521, 502
(1999).

8. S. A. Colgate, Astrophys. J. 150, 163 (1967).

9. M. B. Davies, A. King, S. Rosswog, et al., Astrophys.
J. 579, L63 (2002).

10. V. I. Dokuchaev, Usp. Fiz. Nauk 161, 1 (1991) [Sov.
Phys. Usp. 34, 447 (1991)].

11. V. I. Dokuchaev, Yu. N. Eroshenko, and L.M. Ozernoy,
Bull. Am. Astron. Soc. 29, 848 (1997).

12. V. I. Dokuchaev, Yu. N. Eroshenko, and L.M. Ozernoy,
Astrophys. J. 502, 192 (1998).

13. V. I. Dokuchaev, Yu. N. Eroshenko, and S. G. Rubin,
arXiv:0709.0070v2 [astro-ph] (2007).

14. Y. N. Efremov, B. G. Elmegreen, and P. W. Hodge,
Astrophys. J. 501, L163 (1998).

15. P. Erwin and D. Gadotti, arXiv:1002.1461v1 [astroph.
CO] (2010).

16. S. Graziani, D. Lamb, and J. Quashnock, AIP Conf.
Proc. 428, 161 (1998).

17. L. P. Grishchuk, V. M. Lipunov, K. A. Postnov, et al.,
Usp. Fiz. Nauk 171, 3 (2001) [Phys. Usp. 44, 1
(2001)].

18. W. Hillebrandt, P. Hoflich, P. Kafka, et al., Astron.
Astrophys. 180, L20 (1987).

19. J. G. Hills, Astrophys. J. 267, 322 (1983).

20. V. S. Imshennik, Space Sci. Rev. 74, 325 (1995).

21. V. S. Imshennik and D. K. Nadyozhin, Usp. Fiz. Nauk
156, 561 (1988).

22. V. S. Imshennik and D. K. Nadyozhin, Pis'ma Astron.
Zh. 18, 195 (1992) [Sov. Astron. Lett. 18, 79 (1992)].

23. V. S. Imshennik, O. G. Ryazhskaya, Pis'ma Astron.
Zh. 30, 17 (2004) [Astron. Lett. 30, 14 (2004)].

24. A. King, P. T. O'Brien,M. R. Good, et al., Astrophys.
J. 630, L113 (2005).

25. A. G. Kuranov, S. B. Popov, and K. A. Postnov, Mon.
Not. R. Astron. Soc. 395, 2087 (2009).

26. A. V. Kuznetsov, arXiv:astro-ph/0111014v1 (2001).

27. V. Lipunov and E. Gorbovskoy, Astrophys. J. 665,
L97 (2007).

28. G. V. Lipunova, E. S. Gorbovskoy, A. I. Bogomazov,
et al.,Mon. Not. R. Astron. Soc. 397, 1695 (2009).

29. A. MacFadyen and S. E.Woosley, Astrophys. J. 524,
262 (1999).

30. C.Meegan, V. Connaughton, G. Fishman, et al., IAU
Circ. 6518, 1 (1996).

31. H. Mouri and Y. Taniguchi, Astrophys. J. 566, L17
(2002).

32. M. T.Murphy, J.K. Webb, and I. S. Heng,Mon.Not.
R. Astron. Soc. 316, 657 (2000).

33. B. Paczynski, Astrophys. J. 494, L45 (1998).

34. K. A. Postnov and A. G. Kuranov, Mon. Not. R. Astron.
Soc. 384, 1393 (2008).

35. G. D. Quinlan and S. L. Shapiro, Astrophys. J. 321,
199 (1987).

36. M.Ruffert andH.-Th. Janka, Astron. Astrophys. 338,
535 (1998).

37. A. De Rujula, Phys. Lett. B 193, 514 (1987).

38. R. H. Sanders, Astrophys. J. 162, 791 (1970).

39. S. L. Shapiro and S. A. Teukolsky, Astrophys. J. 307,
575 (1986).
40. L. Stella and L. Treves, Astron. Astrophys. 185, L5
(1987).

41. Ya. Yu. Tikhomirova and B. E. Stern, Pis'ma Astron.
Zh. 31, 323 (2005) [Astron. Lett. 31, 291 (2005)].

42. Ya. B. Zel'dovich and M. A. Podurets, Astron. Zh. 42,
963 (1965) [Sov. Astron. 9, 742 (1965)].

\end{document}